\documentclass[5p,times,11pt]{elsarticle}

\usepackage{xcolor}
\usepackage[utf8]{inputenc}
\usepackage{lineno,hyperref}
\usepackage{amsmath}
\usepackage{dontIndentFirst}
\usepackage{comment}
\usepackage[english]{babel}

\journal{arXiv.org}

\begin{document}

\begin{frontmatter}

\title{Adjoint Monte Carlo calculation of charged plasma particle flux to wall}
\author{Simppa Äkäslompolo}
\address{Aalto University, Finland}
\ead{simppa.akaslompolo@alumni.aalto.fi}

\begin{abstract}
This manuscript describes an adjoint/reverse Monte Carlo method to calculate the flux of charged plasma particles to the wall of e.g. a tokamak. Two applications  are described: a fusion product activation probe and a neutral beam injection prompt loss measurement with a fast ion loss diagnostic. In both cases, the collisions of the particles with the background plasma can be omitted.
\end{abstract}

\end{frontmatter}

\section{Introduction}
In the research of magnetically confined fusion, there is a need to study the behaviour of minority particles in the plasma. A typical example is the calculation of the density distribution of slowing down fusion alpha particles in a tokamak reactor. This characterises the ions confined in the plasma. Another interesting feature are the ions that escape the plasma and hit the walls. The escaping ions can damage the wall, but on the other hand, the plasma can be studied by measuring the escaping ion flux.

The usual tool for this kind of studies is a Monte Carlo particle following code that draws an ensemble of markers from a source distribution, such as fusion reactivity. The trajectory of each marker is then followed according to the equations of motion derived from the Lorentz force, while applying a stochastic diffusion operator to account for collisions of the marker with the background plasma. The author is a member of the group using and developing the ASCOT code~\cite{ascot4ref}, so it is used as the model case. Other similar tools include at least OFMC~\cite{ofmc:JPSJ1981}, SPIRAL~\cite{Kramer13_spiral_reference} SPOT~\cite{Schneider05_SPOT_alphas_current_hole} and LOCUST-GPU~\cite{Akers12_LOCUST-GPU}. These codes work well for the calculation of the distribution function, and also for the escaping flux to the large features of the wall. However, if we are interested in the flux to a small detail of the wall, the \emph{target}, the calculation quickly becomes highly inefficient: from the large initial ensemble only a vanishing fraction hits the target.

The adjoint Monte Carlo integration is based on the idea of swapping the roles of the relatively large \emph{sources} and tiny \emph{targets} in the calculation. %

\section{Proposed adjoint calculation method}
\label{sec:adjMethod}

In our case the source $S$ is a source of particles within the plasma, such as fusion reactivity in activation probe simulations\cite{akaslompolo_ecpd2015_pos}. The target could be any small surface of the wall where the flux of particles is of interest. Since the target is small and the source is large, very few markers will find their way to the target, so most of them are ``wasted''. In the adjoint method, however, markers start backward in time from the target and are much more likely to reach the source.

The idea of the method is to check how much flux (how many particles per second) arrives at the target through all possible final states of the particles that have reached the target. Markers are then followed backwards in time from the final states and the sum (integral) over all the source that would have contributed to the final states is gathered. The trick is to only follow particles backwards in time until they have experienced their \emph{last} collision before hitting the target.  The method could be called \emph{final collision integral}.

All possible final states are included by forming a Mon\-te Carlo integral of them. This is the basis for the initialisation. In other words, $N$ markers are launched from the wall element so that each one carries the weight equivalent of $\left(A_W\cdot\Delta v\cdot \Omega_W\right)/N$, where $A_W$ is the surface area of the target, $\Delta v=v_{\max}-v_{\min}$ is the velocity range of the final state and $\mathrm{d}\Omega_W$ is the solid angle of the possible final velocity directions. For a planar target, the final velocity distribution is isotropic in the half-space open to the plasma ($\Omega_W=2\pi$), and the final locations are uniformly distributed on the surface of the target. (Additionally, importance sampling could be applied, but for the sake of simplicity identical weights are used here.)

In the time-reversed calculation, the full gyromotion of the particles is followed without collisions. At each time step, a certain phase space volume's worth ($\mathrm{d}V\cdot\mathrm{d}\Omega\cdot\mathrm{d}v$) of source $S$ is added to the flux integral. Naturally, particles on the same trajectory with exactly the velocity of the time-reversed marker would hit the target, but particles with nearly same location and nearly the same velocity would reach the target as well. The above vague statement is next refined for use in calculating the phase space volume.

\begin{figure}
  \centering
  \includegraphics[width=\linewidth]{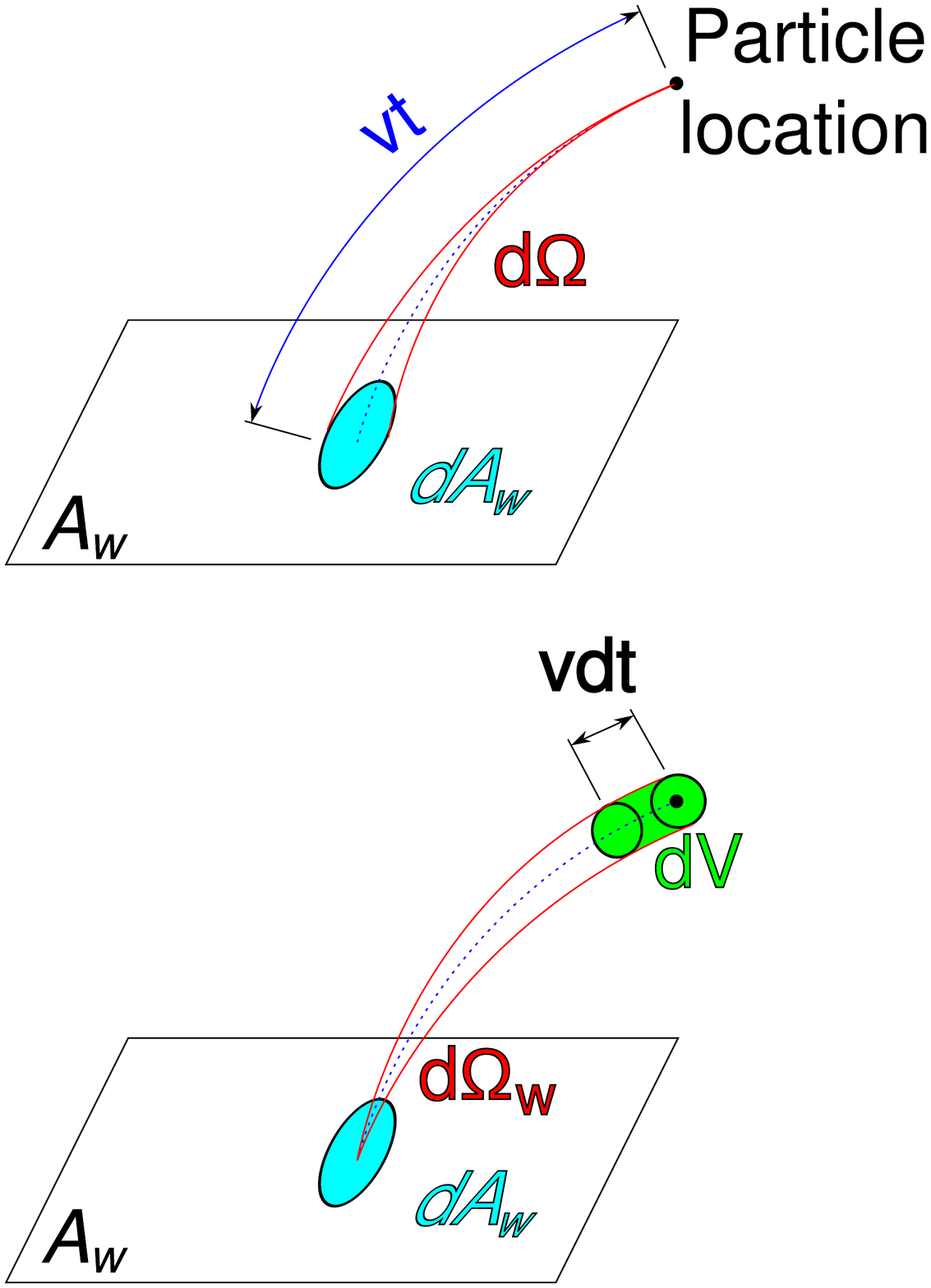}
  \caption{Basis for reverse in time integration}
  \label{fig:rit_scheme_1}
\end{figure}

Figure \ref{fig:rit_scheme_1} illustrates how the phase space volume element $\mathrm{d}V\cdot\mathrm{d}\Omega\cdot\mathrm{d}v$ are related to the Monte Carlo integral weights: 
\begin{equation}
\mathrm{d}A_W\cdot\mathrm{d}v\cdot \mathrm{d}\Omega_W=\frac{A_W\cdot\Delta v\cdot \Omega_W}{N}\label{eq:weights}
\end{equation}
The key idea behind the concept is considering that the target wall surface element $\mathrm{d}A_W$ is visible in the solid angle $\mathrm{d}\Omega$ at the particle location. On the other hand, the solid angle element $\mathrm{d}\Omega_W$ and the (reverse) time step $\mathrm{d}t$ define a cylinder-like volume $\mathrm{d}V$. The calculation is based on the general solid angle formula $\Omega\equiv A/r^2$. The distance $r=vt$ can be calculated along the particle trajectory instead of a straight line, as Liouville's theorem suggests that a curved path along the particle trajectory will conserve the phase space element. We arrive at the following equations:
\begin{eqnarray}
\mathrm{d}\Omega&=&\frac{\mathrm{d}A_W}{\color{green}(vt)^2}\\
\mathrm{d}V&=&v\mathrm{d}t\cdot\mathrm{d}\Omega_W\cdot{\color{green}(vt)^2},
\end{eqnarray}
 where the {\color{green}$r^2=(vt)^2$} part actually cancels out when the phase space volume element components $\mathrm{d}v$, $\mathrm{d}\Omega$ and $\mathrm{d}V$ are multiplied.

The modelling of the collisions of fast particles backward in time is at least non-trivial, if not impossible. This is suggested by the forward collisions producing a Max\-wel\-lian distribution regardless of the initial distribution. Such collisions increase the entropy of the system or in other words remove information. Any reverse calculation would need to supply information to the system, which seems impossible.

This problem is averted either by simply ignoring the collisions, if possible, or following only collisionless orbits backwards in time. The missing collisions are compensated for by including all the collisions in a forward Monte Carlo calculation that calculates the source $S$. The adjoint integration phase includes only the path the particles take after their final collision before hitting the target. This is achieved by weighting each contribution to the integral with the probability $C$ of the marker travelling to the target from the source location $\bf R$ without a collision.

The integral of particle flux to the wall elements is then:

\begin{eqnarray}
  \label{eq:flux}
  \mathrm{F}&=&\iiint\displaylimits_{A_W \Delta v \Omega_W}\int_{t_\mathrm{prt}}S({\bf R,v})C({\bf R,v})\mathrm{d}V\mathrm{d}\Omega\mathrm{d}v\\
&=&\iiint\displaylimits_{A_W \Delta v \Omega_W}\int_{t_\mathrm{prt}}S({\bf R,v})C({\bf R,v})v\mathrm{d}t\mathrm{d}\Omega_W\mathrm{d}v\mathrm{d}A_W,\label{eq:generalFlux}%
\end{eqnarray}
where ${\bf v}$ depicts the velocity vector of the time-reversed marker and $t_\mathrm{prt}$ is the life time of the marker.

%
%

\begin{comment}
\end{comment}

\section{Application for fusion product simulation}

A fusion product activation probe~\cite{actProbe} is a diagnostic probe that can measure the number of high energy ($\sim$MeV) fusion products impacting on it. It is inserted near the plasma in a tokamak. Fusion products activate the samples in the probe, and the activity is measured post mortem.

For collisionless fusion products, the method resembles the calculation of detection efficience $\epsilon$ in~\cite{TokamakIonTemperatureAndPoloidalFieldDiagnosticsUsing3MeVprotons}, but has fewer analytic approximations. The method presented in this manuscript was indepently developed.

When simulating the fusion product activation probe in ASDEX Upgrade tokamak, we have made certain approximations: (1) We neglect the collisions of fusion products after birth; in other words $C\equiv 1$. (2) We use an isotropic monoenergetic fusion product source. In reality the particles have an anisotropic and relatively wide birth energy distribution. The simplified birth distribution allows us to: (a) drop the integral over $\Delta v$ in equation~\eqref{eq:generalFlux} and (b) consider the solid angle dependency simply by dividing by $4\pi$. This will result in the following source term formulation:
\begin{equation}
S\equiv\frac{\left<\sigma v\right>n_1n_2}{4\pi}, \label{eq:fusSource}
\end{equation}
where $\left<\sigma v\right>n_1n_2$ is the fusion reactivity in units $1/({\mathrm{m}^{3}\mathrm{s}})$. (In the general case $S$ would be a function of location and velocity. Thus it would need to have the units  $[S]=1/\left(\mathrm{s}\cdot\mathrm{m/s}\cdot\mathrm{m}^3\cdot\mathrm{srad}\right)$.)

This will result in the following integral for the particle flux to the target:
\begin{equation}
  \mathrm{F}=\iint\displaylimits_{A_W\Omega_W}\int_{t_\mathrm{prt}}\underbrace{\frac{\left<\sigma v\right>n_1n_2}{4\pi}}_\mathrm{precalculated}v\mathrm{d}t\,\mathrm{d}\Omega_W\mathrm{d}A_W
\end{equation}
The source term can be separately precalculated into a $(R,z)$ grid. The rest of the integral can also be stored as a function of $R$ and $z$, so it too is precalculated and stored in a grid, as an \emph{adjoint density}. This makes it practical to quickly change the calculated reactivities or adjoint densities while keeping the other quantity intact: the final flux is calculated as a sum over element-wise multiplied reactivity and time-reversed density.

The first tests of the above scheme against experimental results have been reasonably successful~\cite{akaslompolo_ecpd2015_pos}.

\section{Outline of application to NBI prompt losses to a fast ion loss detector or Faraday cups}
\label{sec:prompt}

The the Faraday cups~\cite{:/content/aip/journal/rsi/75/10/10.1063/1.1788876} and the fast ion loss detector (FILD)~\cite{garcia-munoz:053503} are other escaping fast ion diagnostics. The calculation method described for an activation probe would certainly be adaptable for the FILD, and probably also for the Faraday cups, but since the author has no experience working with Faraday cups, this discussion focusses on the FILD.

The FILD probe is usually sensitive to neutral beam injected ion prompt losses. This means that the fast deu\-te\-rons ($\sim$100\,keV) hit the walls (or the FILD probe) very quickly after being ionized by the plasma. They have little to no time to collide with the background plasma. It would probably be acceptable to study these losses even neglecting collisions. Forward Monte Carlo studies have been made~\cite{asuntaSimsAUGRMP}, but the simulations require huge resources and still cannot provide good statistics in the probe or reproduce the detailed probe geometry. Using the adjoint Monte Carlo method could provide an efficient  calculation of the absolutely calibrated flux of prompt ions to the probe, possibly with detailed geometry.  Time-reversed calculations are already performed routinely with the GOURDON code~\cite{GOURDON-code} to study the final orbits of the measured ions, but this is not equivalent to an adjoint calculation.

When adapting the calculation, the only significant modification from the method described for the activation probe is changing the source term $S$ from \eqref{eq:fusSource} to describe the NBI ions. Here we present two possible schemes.

At the moment the initial ensemble of markers for the forward Monte Carlo calculation in ASCOT is usually created by calculating the ionization locations with the Monte Carlo method~\cite{Asunta14_BBNBI_reference}. The straightforward method would be to bin the initial ion locations to a multidimensional histogram describing the phase space as a function of, e.g., a convenient subspace of $(R,\phi,z,v_R,v_\phi,v_z)$ (major radius, toroidal angle and vertical coordinate, and the respective velocities) and then use this histogram as $S$.

\begin{figure}
  \centering
  \includegraphics[width=\linewidth]{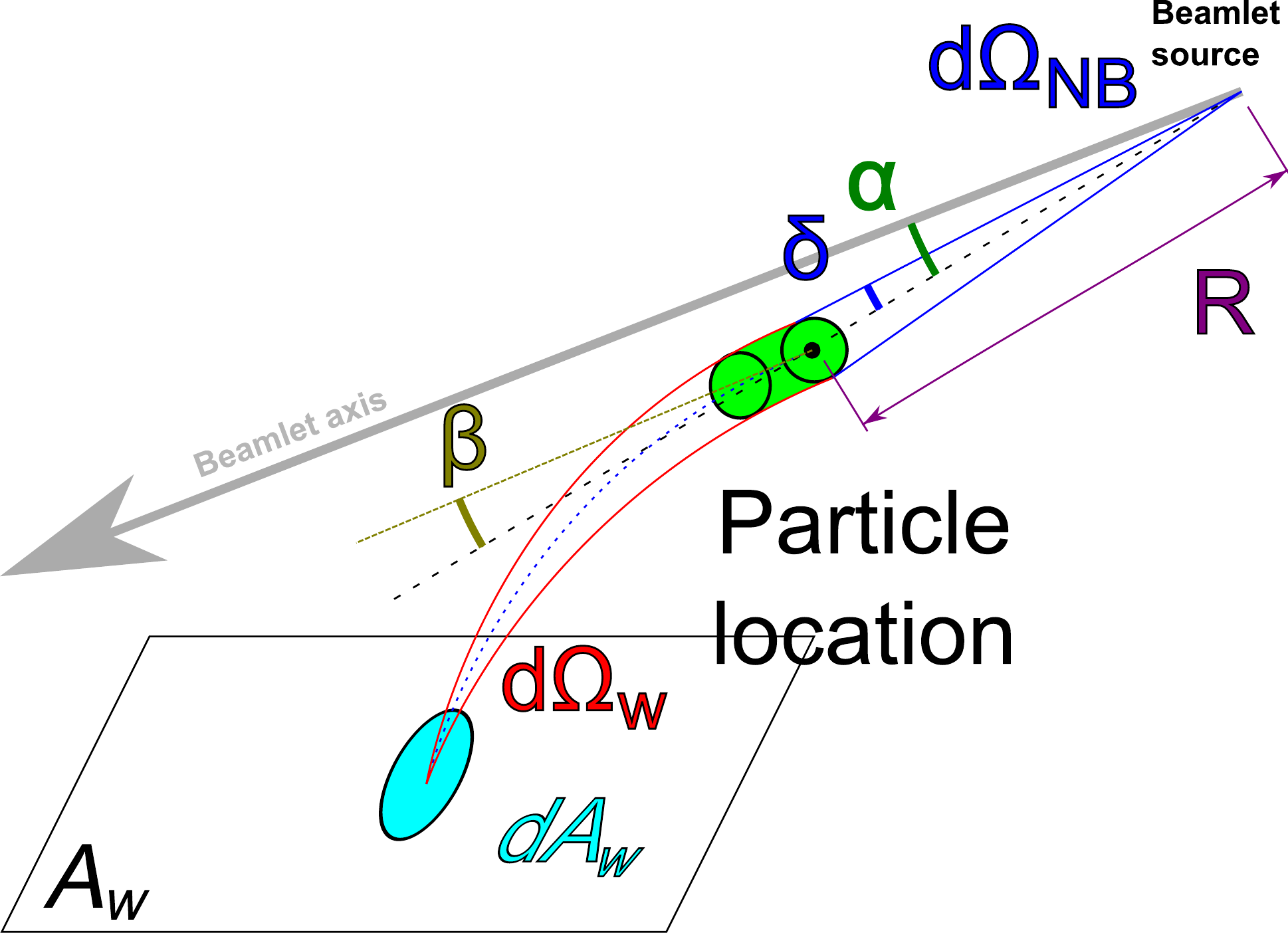}
  \caption{Basis for NBI source from beamlets}
  \label{fig:rit_scheme_nbi}
\end{figure}

Another possibility would be to utilize the geometrical description of the NBI injector model. This is illustrated in figure~\ref{fig:rit_scheme_nbi}. The ionization probability at a given phase space location could possibly be calculated along the way. The beam consists of many beamlets which have a source point, a direction and typically a (bi)gaussian divergence, which means that the beam weakens as a function of angle $\alpha$ shown the figure. The NBI power directed at the given phase space location could then be calculated directly from the NBI model data. However, the beamlet is attenuated in the plasma due to more and more atoms getting ionized. This attenuation could conceivably be tabulated prior to the actual adjoint calculation for quick look-up as a function of distance $R$ from the beamlet source. The solid angle $\mathrm{d}\Omega_\mathrm{NB}$  seen by the beamlet source of the volume element $\mathrm{d}V$ could be calculated as
\begin{equation}
  \label{eq:omegaNB}
\mathrm{d}\Omega_\mathrm{NB}=\mathrm{d}\Omega_W\left(\frac{vt}{R}\right)^2 .
\end{equation}
The ionization probability would be ionizations per solid angle visible from the injector $\mathrm{d}\Omega_\mathrm{NB}$, length along beamlets $v\mathrm{d}t$ and per second.
Finally, the effective solid angle used in the integration should be the ``intersection'' of $\mathrm{d}\Omega_\mathrm{NB}$ and $\mathrm{d}\Omega$.  In other words, the angle $\beta$ between the velocities of the neutral and the time-reversed marker must be sufficiently small; otherwise, the particle would not hit $\mathrm{d}A_W$.

The first attempt at prompt loss calculation should probably be made with the binning method, as the latter method seems unnecessarily complicated.

\section{Summary and discussion}

We have described a method to perform adjoint Monte Carlo calculation of fusion products. This method has been already tested with promising results \cite{akaslompolo_ecpd2015_pos}. We have also outlined a method for neutral beam injection (NBI) prompt losses, but this has not been implemented or tested yet.

\paragraph{Future work} would be to test the NBI prompt loss method of section~\ref{sec:prompt}. The really big deal, however, would be to come up with a way to do collisional adjoint calculations. This would require either time-reversed collision operators, or a method to calculate some sort of source term for collided particle flux.  The section \ref{sec:adjMethod} has been written with such a source term in mind. Conceivably, the flux of particles within the plasma could be calculated with a forward Monte Carlo calculation that includes the collisions. This flux could then be ``collided'' against a suitable differential (Coulomb) collision/scattering cross section to produce the sought-after source term. The details of such a method remain future work, possibly to be published in a future version of this arXiv.org manuscript.

\section*{Acknowledgements}
Discussions with Risto Vanhanen and Pertti Aarnio were very helpful. Seppo Sipil\"a and Otto Asunta proof-read the manuscript.

This work was partially funded by the Academy of Finland project No. 259675. This work  has received funding from Tekes – the Finnish Funding Agency for Innovation under the FinnFusion Consortium. 

\section*{References}

\bibliography{../bibfile}

\end{document}